# Design of resonators using materials with negative refractive index

Petra Schmidt[1], Ilya Grigorenko[3,4], and A.F.J. Levi[1,2]

[1]*Department of Physics and Astronomy*

*University of Southern California*

*3620 S. McClintock Avenue, SSC 211A*

*Los Angeles, CA 90089-0484*

[2]*Department of Electrical Engineering*

*University of Southern California3620 S. Vermont Avenue, KAP 132*

*Los Angeles, CA 90089-2533*

[3]*Theoretical Division T-11, Los Alamos National*

*Laboratory, Los Alamos, NM 87545*

[4]*Center for Nonlinear Studies, Los Alamos National*

*Laboratory, Los Alamos, NM 87545*

**Abstract**

By optimizing the design we show that inhomogeneous electromagnetic resonators with almost uniform field intensity and up to twice the energy density of conventional structures are possible by exploiting the properties of negative refractive index materials. We demonstrate that using negative refractive index materials it is possible to make full width at half maximum (FWHM) of the transmission coefficient independent from cavity length *L*.

The linear response of a material to the presence of an electromagnetic wave at frequency ω is characterized by the permittivity ε(ω) and permeability μ(ω) of the medium. Materials that exhibit both negative permittivity and permeability at frequency ω have negative refractive index at that frequency. A consequence is that the phase velocity of electromagnetic waves propagating in negative index material is directed against the flow of energy [1]. While naturally occurring materials do not exhibit high frequency magnetism, Pendry showed how the inductive response of non-magnetic metal structures could be used to obtain negative permeability at high frequency and hence negative refractive index material with unique properties [2]. RF experiments confirmed these ideas [3]. Some recent experiments with nanofabricated metal structures suggest that negative refractive index may be also obtained at visible frequencies [4,5].

The ability to create materials with negative refractive index is an additional degree of freedom for the design of RF and optical devices. In this paper we show how negative index material can be used to control the electric field intensity profile in the cavity of an electromagnetic resonator. Although currently available negative index materials are very lossy, researchers spend considerable efforts to improve this characteristic. For example, Popov and Shalaev recently suggested a promising way to compensate losses using optical parametric amplification [6]. Throughout the paper we assume negative index materials with negligible losses.

The boundary conditions for a monochromatic linearly polarized electromagnetic wave incident normal to a planar interface between positive refractive index $n_1 > 1$ ($\mu_2(\omega) > 0$, $\varepsilon_1(\omega) > 0$) and a region with negative refractive index $n_2 < -1$ ($\mu_2(\omega) < 0$, $\varepsilon_2(\omega) < 0$) are

$$E_x\big|_{z=z_0-\delta} = E_x\big|_{z=z_0+\delta} \tag{1}$$

$$\frac{1}{\mu_1(\omega)}\frac{\partial E_x}{\partial z}\bigg|_{z=z_0-\delta} = \frac{1}{\mu_2(\omega)}\frac{\partial E_x}{\partial z}\bigg|_{z=z_0+\delta} \tag{2}$$

where $z$ is in the normal direction, $z_0$ is the position of the interface, $\delta$ is a small positive $z$-directed shift in position, and the electric field oscillating at frequency ω is linearly polarized in the $x$-direction. Eqn. (2) arises from continuity of magnetic field parallel to the interface and Maxwell's equation for the curl of the electric field. Note, that Eqn. (2) makes possible sign change for the derivative of the field's x-component at the interface of normal-negative index materal. It results in much better control over the field in the resonator by engineered spatially inhomogeneous layered optical medium, that simply was not possible using usual materials.

To illustrate optimal design approach let us consider a resonator created using two high-reflectivity Bragg mirrors, each consisting of 5 identical layer pairs in a planar slab geometry. Individual dielectric layers have thickness $\lambda_0/(4n)$ where $n$ is the refractive index of the dielectric and $\lambda_0$ is the resonant wavelength. The length $L$ separating the two mirrors defines the extent of the resonator cavity. For the purpose of illustration we choose $\lambda_0 = 980$ nm, dielectric pairs consisting of air ($n = 1$) and a material with $n = 2$, and a cavity length $L = \lambda_0/n$ with $n = 2$. The quasi-stationary electric field in the resonator is found using the propagation matrix method [7]. Figure 1(a) shows the refractive index profile normal to the dielectric layer structure. Figure 1(b) shows the corresponding calculated electric field intensity profile for linearly polarized electromagnetic radiation of wavelength $\lambda_0 = 980$ nm and amplitude $|\mathbf{E}_0|$ incident normal to the dielectric layer structure from the left. The electric field intensity plotted in the figure is normalized to the incident intensity. The peak intensity in the cavity region is approximately $I_p = 1024$ and, due to the sinusoidal behavior of $|\mathbf{E}(z)|^2$, the average value of the electromagnetic energy density in the cavity is half this value.

Resonant structures similar to that illustrated in Fig. 1 are typically used to store electromagnetic energy or to enhance interaction of electromagnetic radiation with matter.

Hence, an appropriate design goal is maximizing electromagnetic energy density in the cavity or, equivalently, maximizing

$$W = \int_{z_L}^{z_R} |\mathbf{E}(z)|^2 dz, \qquad (3)$$

where $z_L$ and $z_R$ are the left and right boundaries of the resonant cavity respectively $z_R$ $z_L$.

In an ad-hoc approach one might consider the cavity to be made up of dielectric pairs [8] consisting of alternating layers of $n = 2$ and $n = -2$ material, each of thickness $d$. Figure 2(a) shows the refractive index profile for the case when there are six layer pairs in the cavity and $d = \lambda_0/(12|n|)$. Figure 2(b) shows the result of calculating electromagnetic field intensity profile. Peak intensity in the cavity is still $I_p = 1024$, but $|\mathbf{E}(z)|^2$ is no longer sinusoidal, and the value of $W$ is enhanced by a factor of $\xi = 1.826$. As indicated in the inset of Fig. 2(a), for three or more layer pairs, increasing the number of layer pairs that fill the cavity of length $L = \lambda_0/|n|$ has the effect of increasing the energy density. With increasing layer pairs the enhancement factor asymptotically approaches $\xi = 2$. The standard deviation in cavity field intensity decreases by almost a factor of 5 from a value of 362 for the conventional case illustrated in Fig. 1(b) to 77 for the result shown in Fig. 2(b).

Optimal design of the cavity is achieved using Richard Brent's "principal axis" optimization method [9,10] for a scalar function of $N$ variables. In this work we choose $N = 6$ corresponding to six layer pairs in the cavity. Each layer pair has thickness $\lambda_0/(6|n|)$. The optimized parameters are the thicknesses $0 \leq d_i$ ($i=1,\ldots,6$) $< \lambda_0/(6|n|)$ of material with $n = 2$. The thickness of the pair layer of material with $n = -2$ is $\lambda_0/(6|n|) - d_i$. Figure 3(a) shows the results of optimization with the objective of maximizing $W$. Compared to the ad-hoc design, field intensity in the cavity region is more uniform and the value of the enhancement factor increases by 0.117 to $\xi = 1.943$. The standard deviation in cavity field intensity decreases to 29.

Applying the same optimization procedure but allowing only positive layer pair material refractive index values $n = 1, 2$ results in an index profile identical to that shown in Fig. 1. It is the negative refractive index degree of freedom that allows access to the remarkable results illustrated in Fig. 2 and Fig. 3. Additional degrees of freedom, for example allowing the refractive indices of individual layers in the resonator to vary from the symmetric choice $n = 2, -2$, make the optimal design objectives more accessible.

The calculated transmission coefficient for each structure illustrated in Figs. 1-3 as a function of the wavelength $\lambda$ is shown in Fig. 4. Each can be fit to a Lorenzian. For the structure shown in Fig. 1 the full width at half maximum (FWHM) is 0.203 nm with the chi-square estimator $\chi^2 = 1.73 \times 10^{-6}$. For the structures shown in Fig. 2 and Fig. 3 FWHM is 0.610 nm with $\chi^2 = 1.30 \times 10^{-5}$.

Conventional resonator designs of the type illustrated in Fig 1(a) have a transmission spectrum peak with FWHM proportional to inverse cavity length ($1/L$). For fixed distributed Bragg-reflector (DBR) design, stored electromagnetic energy in the cavity can only be enhanced by increasing $L$ and decreasing FWHM. Because temporal response is proportional to $e^{-t/\tau}$, the characteristic lifetime $\tau$, proportional to the inverse FWHM, increases.

In Fig 5 (a) we show a contour plot for the transmission coefficient as a function of wavelength $\lambda$ and the cavity length $L$ for a conventional resonator, which consists of 5 DBR pairs, alternating layer pairs of $n = 2$ and $n = 1$ material (the design is similar to what is shown in Fig. 1(a)). Note, we plot the transmission coefficient using the $\log_{10}$ scale. The contour plot exhibits a complex structure that is originated from the phase accumulation in the cavity.

The situation is quite different for resonator designs of the type illustrated in Fig. 2 and Fig. 3. They have FWHM that is *independent* of L and FWHM only depends on reflectivity of the DBRs. To illustrate this, in Fig 5 (b) we plot for the transmission coefficient as a function of wavelength λ and the cavity length L for a resonator, which consists of 5 DBR pairs, alternating layer pairs of $n = 2$ and $n = 1$ material, the central cavity region is consisting of alternating layer pairs of $n = 2$ and $n = -2$ material, similar to the design shown in Fig. 2(a).

It is therefore possible to increase resonator device design space by separating τ from electromagnetic energy stored in the cavity. The physical reason for this is phase compensation of an electromagnetic wave propagating through the positive and negative refractive index media in the cavity.

Sensitivity analysis for the conventional uniform design with positive refractive index shown in Fig. 1(a) indicates that a 0.01% change of the total width of the resonator L can decrease the stored energy of the electromagnetic field by a factor 10. This result is in a qualitative agreement with the ratio of the characteristic width of the Lorentzian fitted to the calculated transmission coefficient for the structure, and the resonant transmission wavelength $\lambda_0$ = 980 nm. Sensitivity analysis for the optimized structure illustrated in Fig 3(a) was performed by independently perturbing each layer thickness in the cavity using a random number generator with uniform distribution. We observed a multi-modal solution space with either a significant decrease of W or a result comparable to the optimized solution. Using a statistical average over more than 10000 realizations we estimate sensitivity of the designs shown in Fig. 3(a) to be approximately two times higher than the conventional structure illustrated in Fig. 1(a).

In conclusion, in this work we introduce an optimal design approach to obtain certain optical properties of an inhomogeneous media. We demonstrated that the use of layered materials of positive and negative refractive index on the sub-wavelength scale gives rise of unusual optical properties of resonant electromagnetic cavities. For fixed DBR design, it is possible to decouple FWHM from cavity length $L$, control electromagnetic field intensity, and double stored energy density in a resonator cavity. The results of our simulations are also valid for the case of electron scattering in inhomogeneous potential solid-state structures, since electromagnetic wave propagation formally equivalent to the propagation of electron wavefunction. Based on this observation we conclude, that new functionalities can be obtained through design of similar spatial inhomogeneities for quite different systems. The optimal design approach employed in this work was successfully used in other applications to discover novel types of materials with controlled properties [11,12,13].

This work is supported by DARPA and ONR. I.G. acknowledges the support of the National Nuclear Security Administration of the U.S. Department of Energy at Los Alamos National Laboratory under Contract No. DE-AC52-06NA25396.

**Figure captions**

**Figure 1.** (a) Refractive index profile $n(z)$ of a conventional dielectric resonator consisting of two Bragg mirrors and a cavity. Resonant wavelength is $\lambda_0 = 980$ nm. Each Bragg mirror consists of five layer pairs of $n = 1$ and $n = 2$ material, each layer being $\lambda_0/(4n)$ thick. The central cavity region has refractive index $n = 2$ and length $L = \lambda_0/n$. (b) Normalized resonant electric field intensity $|\mathbf{E}(z)|^2$ corresponding to the refractive index profile shown in (a). Field intensity is normalized to the intensity of the electromagnetic wave incident on the structure from the left.

**Figure 2.** (a) Refractive index profile $n(z)$ of an electromagnetic resonator with a cavity containing dielectric pairs consisting of six alternating layer pairs of $n = 2$ and $n = -2$ material, each of thickness $d$. Resonant wavelength is $\lambda_0 = 980$ nm. Inset shows increasing the number of layer pairs that fill the cavity of length $L = \lambda_0/|n|$ has the effect of increasing the energy density. With increasing layer pairs the stored energy enhancement factor asymptotically approaches $\xi = 2$. (b) Normalized resonant electric field intensity $|\mathbf{E}(z)|^2$ corresponding to the refractive index profile shown in (a).

**Figure 3.** (a) Optimal refractive index profile $n(z)$ of a resonator that maximizes electromagnetic field intensity in the cavity. The cavity contains six alternating layer pairs of $n = 2$ and $n = -2$ material. Thicknesses of the $n = -2$ material are $d_1 = 28.48$ nm, $d_2 = 43.42$ nm, $d_3 = 40.35$ nm, $d_4 = 41.21$ nm, $d_5 = 38.58$ nm, and $d_6 = 52.93$ nm. Resonant wavelength is $\lambda_0 = 980$ nm and the cavity has length $L = \lambda_0 / |n|$. (b) The normalized resonant electric field intensity $|\mathbf{E}(z)|^2$ corresponding to the refractive index profile shown in (a).

**Figure 4.** Transmission coefficient as a function of wavelength $\lambda$ for resonator structure illustrated in Fig. 1(a) (solid line) and Figs. 2(a) and 3(a) (dotted line). The cavity with alternating positive and negative refractive index has a larger line width. The results can be fit to Lorentzian functions with FWHM of 0.203 nm and 0.610 nm.

**Figure 5.** (a) Transmission coefficient for a resonator as a function of wavelength $\lambda$ and cavity length $L$ normalized to the resonant wavelength in the cavity, $\lambda_{res}$. Each DBR mirror consists of 5 alternating layer pairs of $n = 2$ and $n = 1$ material with each layer a quarter wavelength thick and a central cavity region of length $L$ with refractive index $n = 2$, similar to Fig. 1(a)). Note the complex structure. The transmission coefficient is plotted using a $\log_{10}$ scale. (b) Transmission coefficient for the structure in (a) but with the central cavity region consisting of alternating layer pairs of $n = 2$ and $n = -2$ material. Note, that unlike in (a), transmission is independent of the cavity length $L$ because of electric field phase cancellation in the cavity.

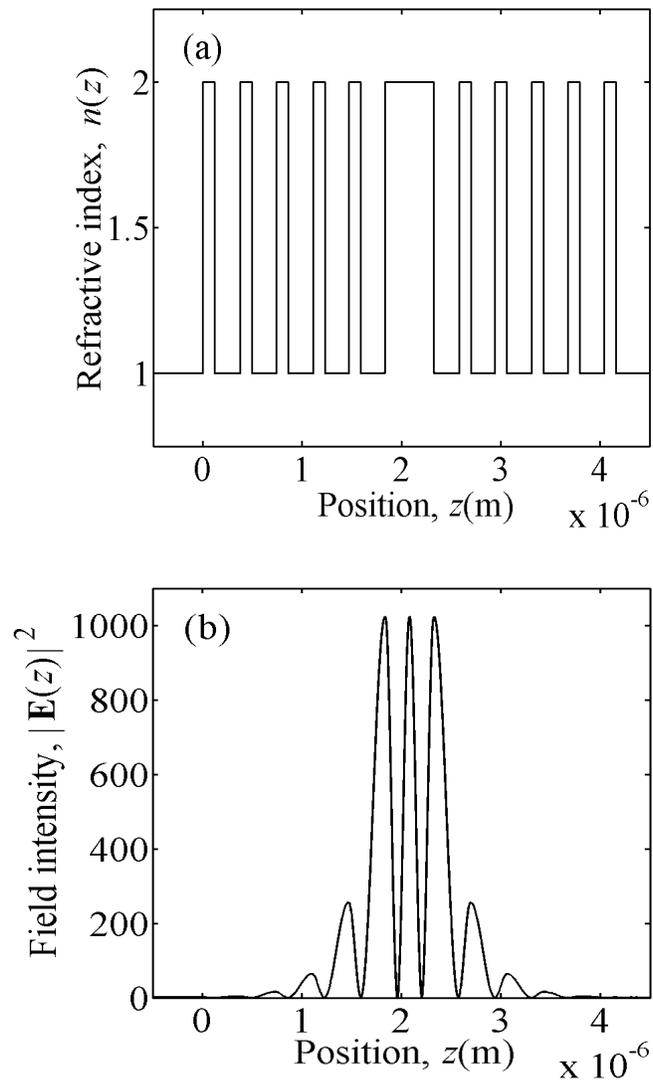

**Figure 1:**

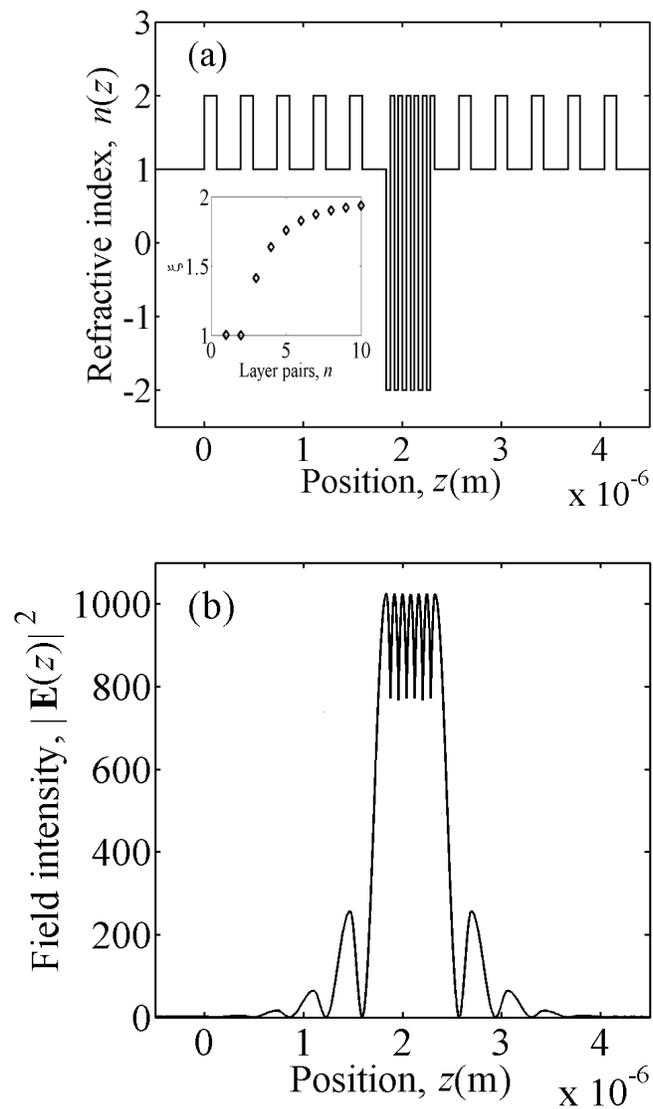

**Figure 2:**

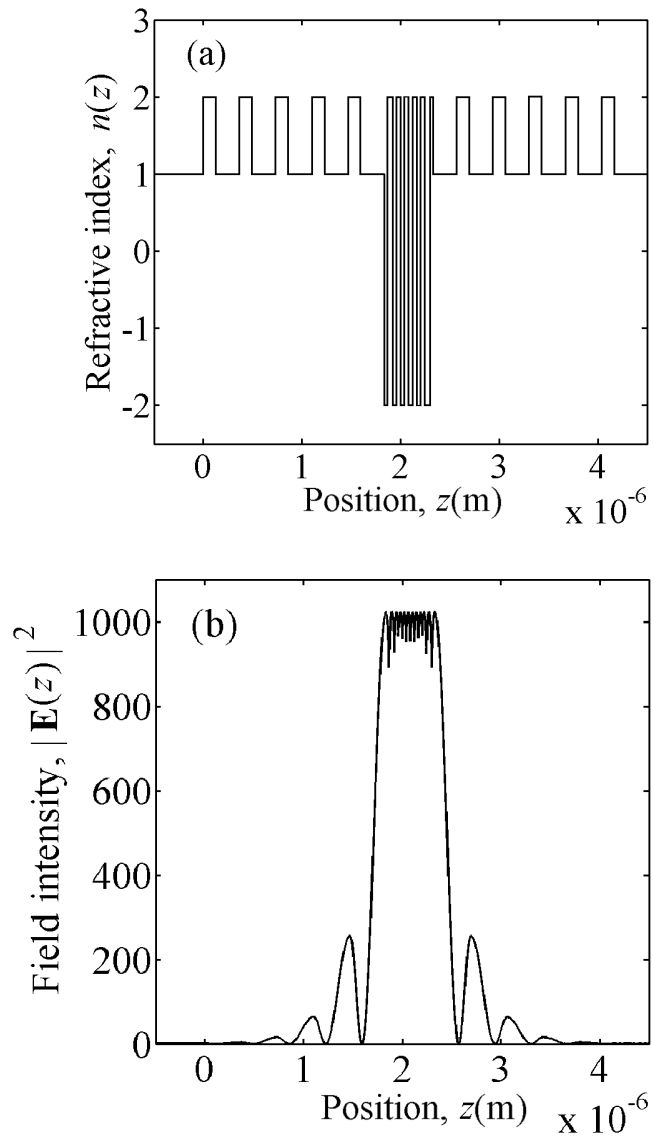

**Figure 3:**

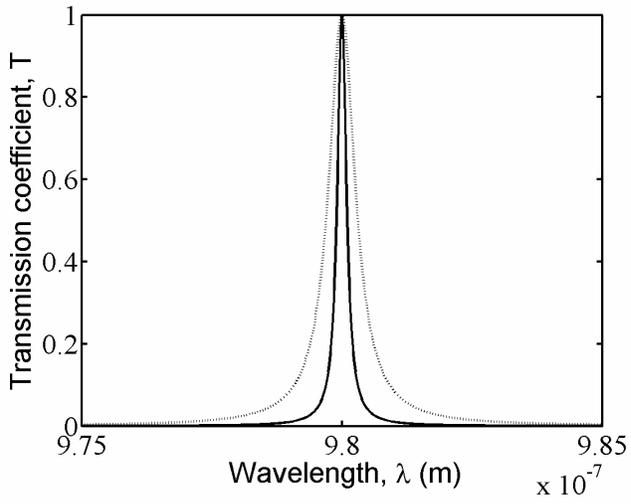

**Figure 4:**

**(a)**

**(b)**

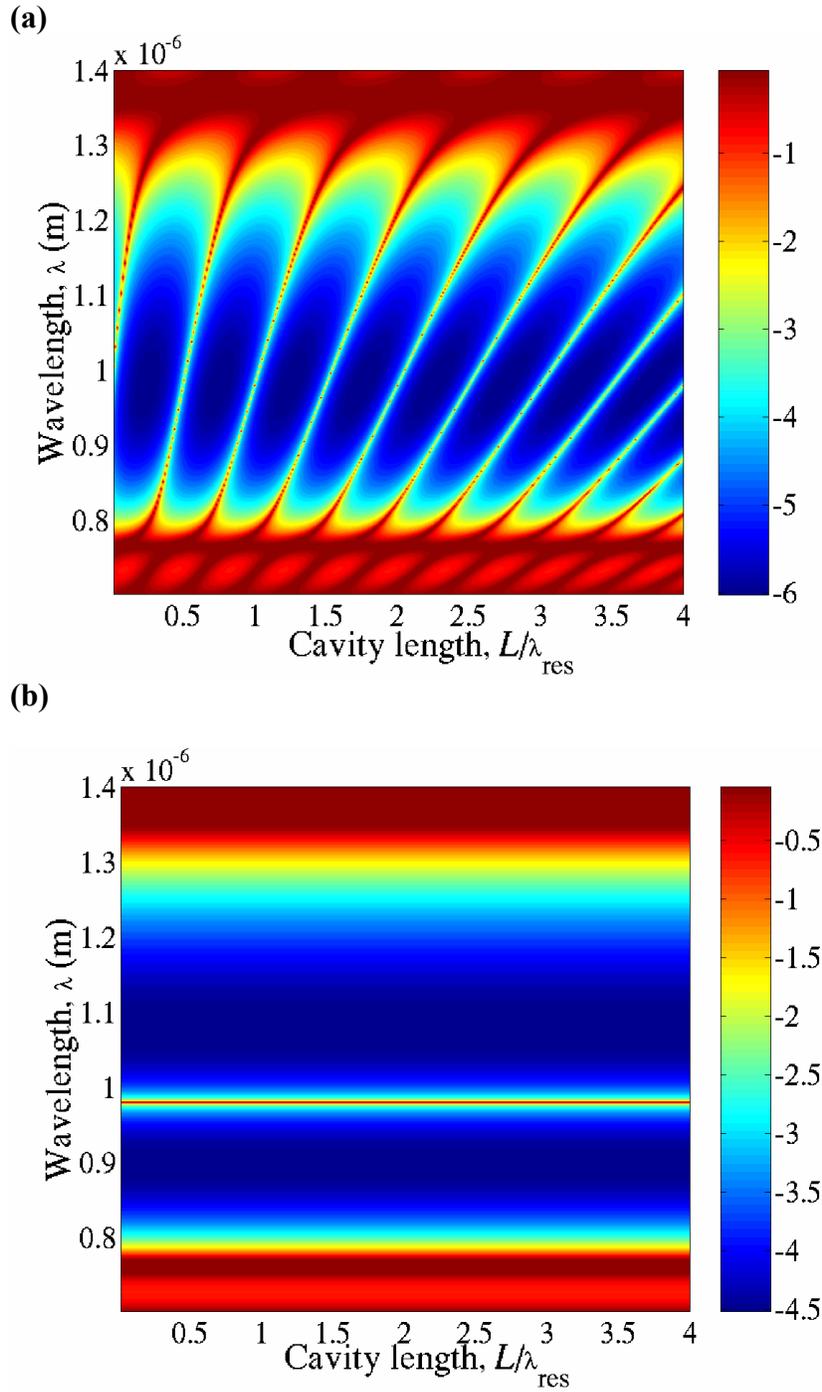

**Figure 5:**